\documentclass{emulateapj}

\newcommand{\expnt}[2]{\ensuremath{#1 \times 10^{#2}}}   
\newcommand{\gsim}{\gtrsim}

\def\micron {\ensuremath{\mu\mbox{{m}}}}
\newcommand{\um}{\micron}

\newcommand{\ks}{\ensuremath{K_{\rm s}}}

\newcommand{\av}{\ensuremath{A_{V}}}

\newcommand{\ujy}{\ensuremath{\mu\mbox{{Jy}}}}

\newcommand{\axp}{1E~2259+586}

\defcitealias{wck06}{WCK06}

\shortauthors{Kaplan et al.}
\shorttitle{A Mid-Infrared Counterpart to \axp}

\begin{document}

\title{A Mid-Infrared Counterpart to the Magnetar \axp}

\author{David~L.~Kaplan\altaffilmark{1,2} and Deepto~Chakrabarty}
\affil{Kavli Institute for Astrophysics and Space Research,
  Massachusetts Institute of Technology, Cambridge, MA 02139}
\email{dlk@space.mit.edu, deepto@space.mit.edu}

\author{Zhongxiang~Wang} 
\affil{Department of Physics, McGill University, Montreal, QC
  H3A~2T8, Canada}  
\email{wangzx@physics.mcgill.ca}

\and

\author{Stefanie~Wachter}
\affil{Spitzer Science Center, California Institute of
 Technology, Pasadena, CA 91125}
\email{wachter@ipac.caltech.edu}

\slugcomment{Astrophysical Journal, in press}

\altaffiltext{1}{Hubble Fellow}
\altaffiltext{2}{Current address: KITP, Kohn Hall, UCSB, Santa
  Barbara, CA 93106-4030.}

\begin{abstract}
We report the discovery of a $4.5\,\um$ counterpart to the anomalous
X-ray pulsar (magnetar) \axp\ with the {\em Spitzer Space Telescope}.
The mid-infrared flux density is $6.3\pm1.0\,\ujy$ at $4.5\,\um$ and
$<$20\,$\ujy$ (at 95\% confidence) at $8\,\um$, or $0.02$\% of the
2--10\,keV X-ray flux (corrected for extinction).  Combining our {\em Spitzer}
measurements with previously published near-infrared data, we show
that the overall infrared emission from \axp\ is qualitatively similar
to that from the magnetar 4U~0142+61.  Therefore, the passive
X-ray-heated dust disk model originally developed for 4U~0142+61
might also apply to \axp.  However, the IR data from this source can
also be fitted by a simple power-law spectrum as might be expected from
magnetospheric emission.
\end{abstract}

\section{Introduction}
\label{sec:intro}
The notion of supernova fallback, where some of the ejecta from a
core-collapse supernova ends up captured by the newly formed neutron
star \citep{chevalier89} and may have sufficient angular momentum to
form a disk \citep*{lwb91}, has been a general prediction of supernova
models \citep[e.g.,][]{ww95}.  Fallback can have profound effects on the final
state of the neutron star, forming disks that can give rise to planets
\citep{ph93,pod93}, and possibly even causing the neutron star to
collapse into a black hole.  Fallback disks should manifest as a
thermal infrared excess \citep*{phn00}, but  previous searches
for such  excesses around neutron stars were unsuccessful \citep[][and
  references therein]{lww04}.

Several years ago, though, we discovered the mid-infrared (mid-IR; here, $4.5$
and $8.0\,\um$) counterpart to the magnetar 4U~0142+61
\citep*[][hereafter \citetalias{wck06}]{wck06}.  The combined
optical/IR spectrum of this magnetar suggests that the optical and IR
data arise from two different spectral components. While the optical
component is demonstrably of magnetospheric origin \citep{km02}, we
showed that the IR component may arise from a passive (i.e., not
accreting, but see \citealt{eeea07}), dust disk irradiated by X-rays
from the magnetar (\citetalias{wck06}; \citealt*{wck08}).  The
inferred spectral shape, while not well constrained, is remarkably
similar to those of protoplanetary disks around young stars
\citep{bscg90}.  The disk's survival lifetime ($\gtrsim 10^6$~yr)
significantly exceeds the pulsar's spin-down age\footnote{While
spin-down is not an accurate measure for the age of a magnetar
\citep{wt06}, no independent age estimate is available for 4U~0142+61
and this age is consistent (if not somewhat longer than) typical ages
of other magnetars.} ($\lesssim 10^5$~yr), consistent with a supernova
fallback origin.

This dust disk model, while intriguing, is not definitive.  In
particular, we have yet to firmly establish whether the mid-IR flux in
4U~0142+61 comes from a disk or from the magnetosphere (via any of a
number of mechanisms, e.g., \citealt*{egl02}; \citealt{ec04};
\citealt{lz04b}; \citealt{hh05}; \citealt{bt07}).  Variability at a
single wavelength or across wavelengths can be a powerful
discriminant, depending on the timescale.  This was illustrated by
\citet{km02}, who found that the pulsed fraction in the optical
exceeded that in soft X-rays, making it impossible for the optical to
result from reprocessed X-rays (also see \citealt{dmh+05}).  In the
infrared the situation is less clear.  The significant variability
seen by \citet{dvk06c} at $2.1\,\um$ with no related change in soft
X-rays provides a stiff test for a disk model, but is not definitive
(see \citealt{wk08}).
The unknown behavior at higher energies, the complicated behavior
across the optical/near-infrared (near-IR), and the lack of pulsations at
$2.1\,\um$ \citep[{$<17$\% pulsed at 90\% confidence};][]{mkk+09} all
complicate the matter.

As we attempt to determine the origin of the mid-IR emission, we must
equally attempt to understand how ubiquitous mid-IR counterparts to
young neutron stars are.  Suggestively, all four of the magnetars with
confirmed quiescent near-IR counterparts have the same
near-IR/X-ray flux ratio of $\approx 10^{-4}$ \citep{dvk05,dvk06d}.
In the disk scenario, this ratio should be determined largely by
geometry, so mid-IR observations of other magnetars down to this
fractional level may help establish whether disks are present, but
similar behavior might also be expected from magnetospheric emission.
Shallow mid-IR searches of three other magnetars were not very
constraining \citep*{wkh07}, but deep a $4.5\,\um$ upper limit for
1E~1048.1$-$5937 following an X-ray flare concluded that it did not
have a mid-IR counterpart similar to that of 4U~0142+61 despite its
similar near-IR/X-ray flux ratio, thus casting some doubt on the disk
interpretation \citep{wbk+08}.  Here, we present the results of a deep
search with the \textit{Spitzer Space Telescope} for a mid-IR
counterpart to the magnetar \axp.

\object[1E 2259+58.6]{\axp} was identified as an X-ray point source
with coherent $7$\,s pulsations in the center of the supernova remnant
 (SNR) CTB~109 \citep{fg81}.  The association leads to a distance of $3.0\pm
0.5$~kpc based on interactions between the SNR and \ion{H}{2} regions
with measured distances \citep*{kuy02}, although \citet{dvk06}
determined a distance to \axp\ of $7.5\pm1.0\,$kpc using the ``red
clump'' method.  \citet{dvk06b} found the column density to \axp\ to
be $N_{\rm H}=(1.1\pm0.3)\times10^{22}\,{\rm cm}^{-2}$, corresponding
to a visual extinction of $\av=6.3\pm 1.8\,$mag, from fitting of X-ray
absorption edges (consistent with earlier measurements;
\citealt{pkw+01}).  The near-IR ($\ks$-band, or $2.1\,\um$)
counterpart to \axp\ was identified by \citet{htvk+01}.  Uniquely
among magnetars, the near-IR flux was observed to vary in concert with the X-ray
flux, with both declining following a major series of X-ray bursts
\citep{tkvkd04}.  
This led to the suggestion  of X-ray heating of a disk
\citep[as in][for a low-mass X-ray binary]{mmo+84}, making this source an especially attractive
target for mid-IR searches.

\begin{figure*}
\centerline{\includegraphics[width=0.5\textwidth]{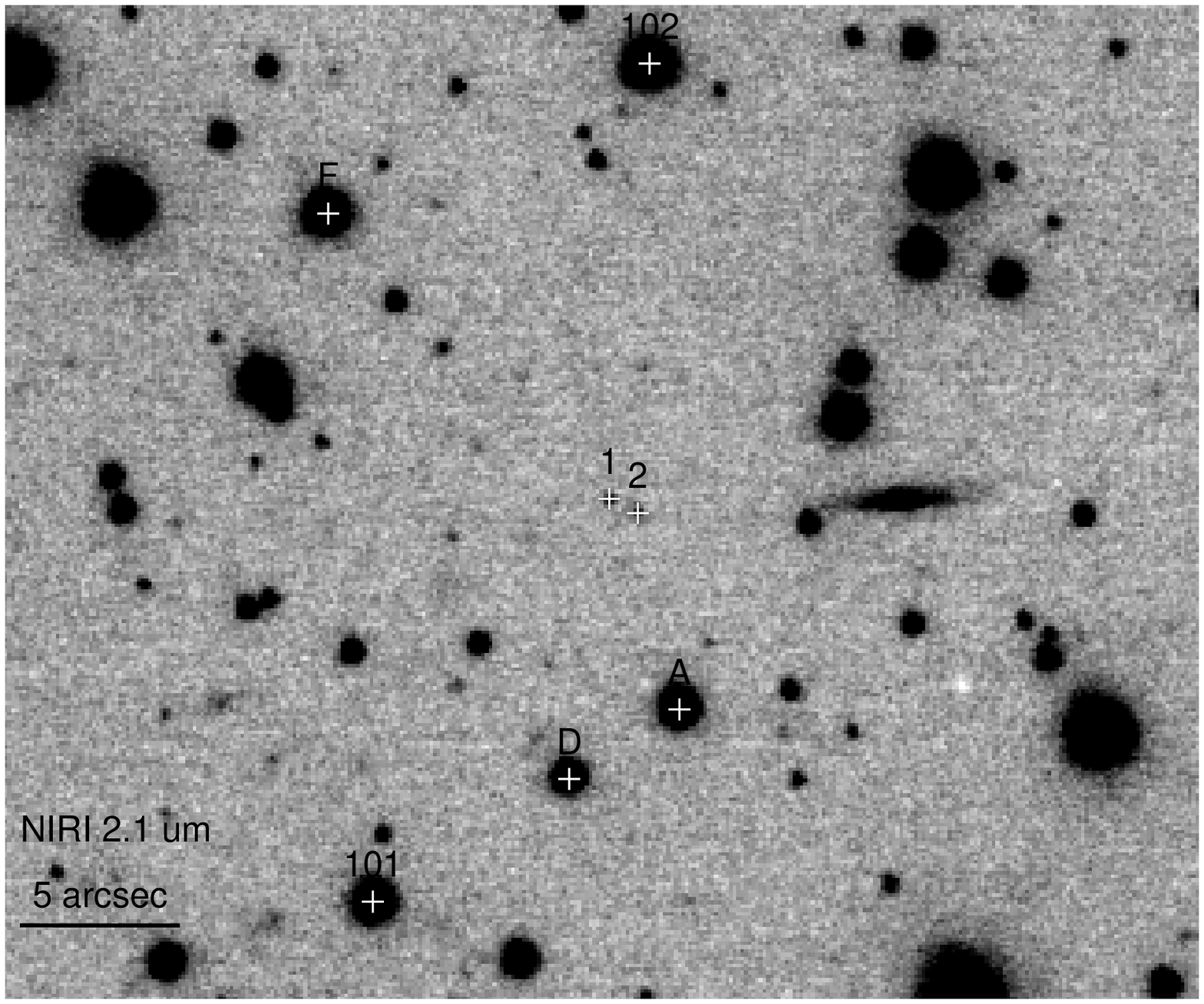}\includegraphics[width=0.5\textwidth]{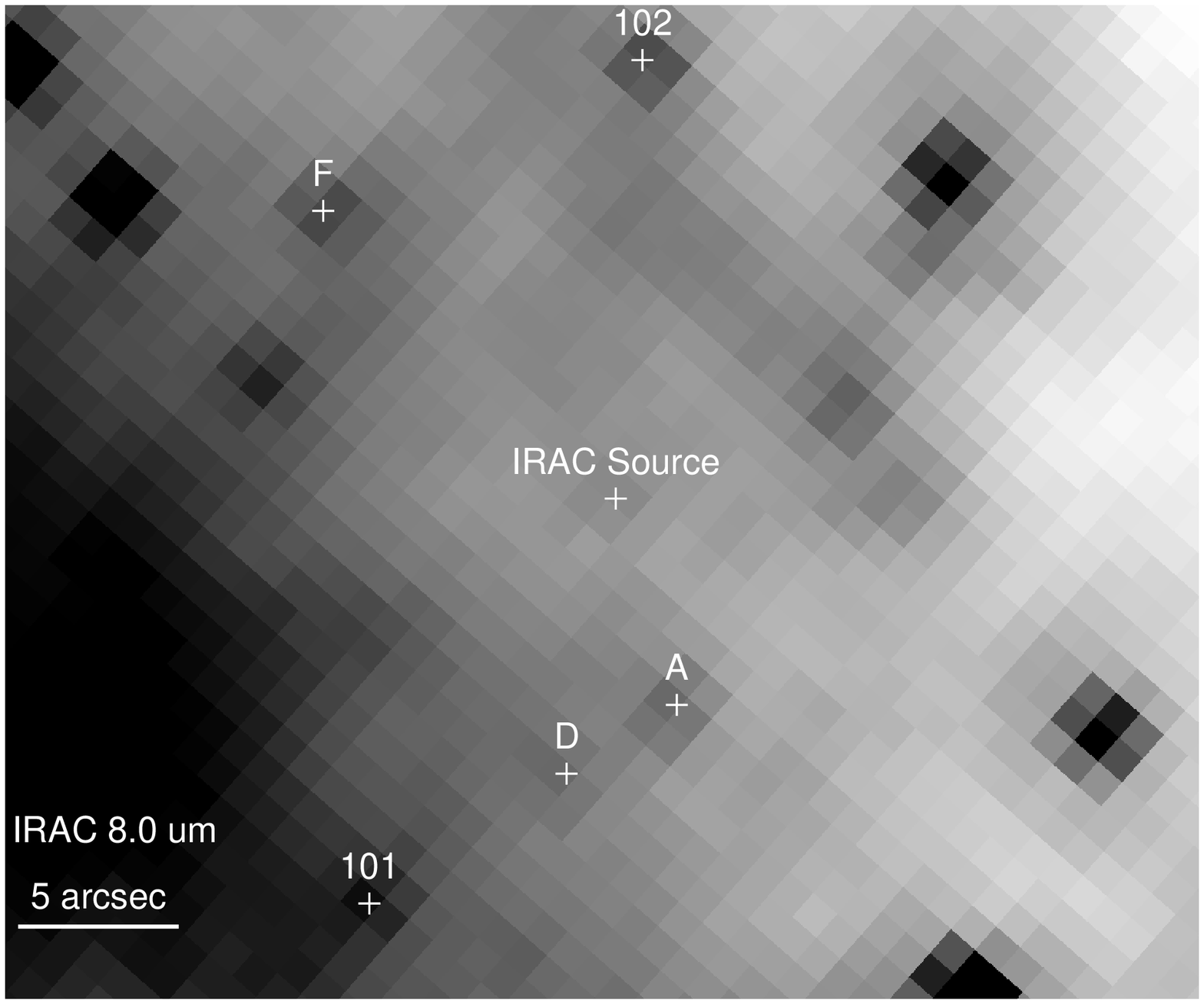}}
\centerline{\includegraphics[width=0.5\textwidth]{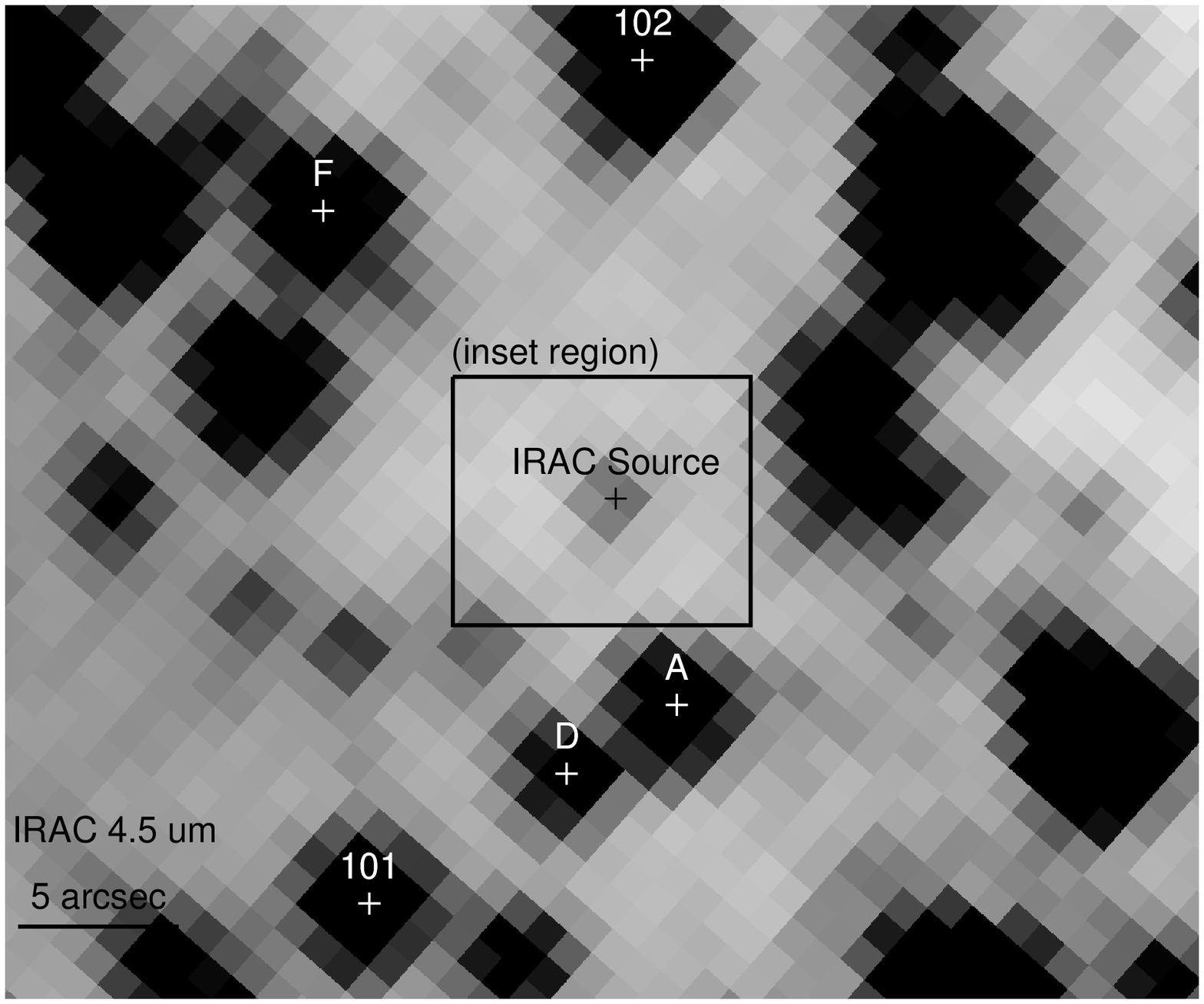}\includegraphics[width=0.5\textwidth]{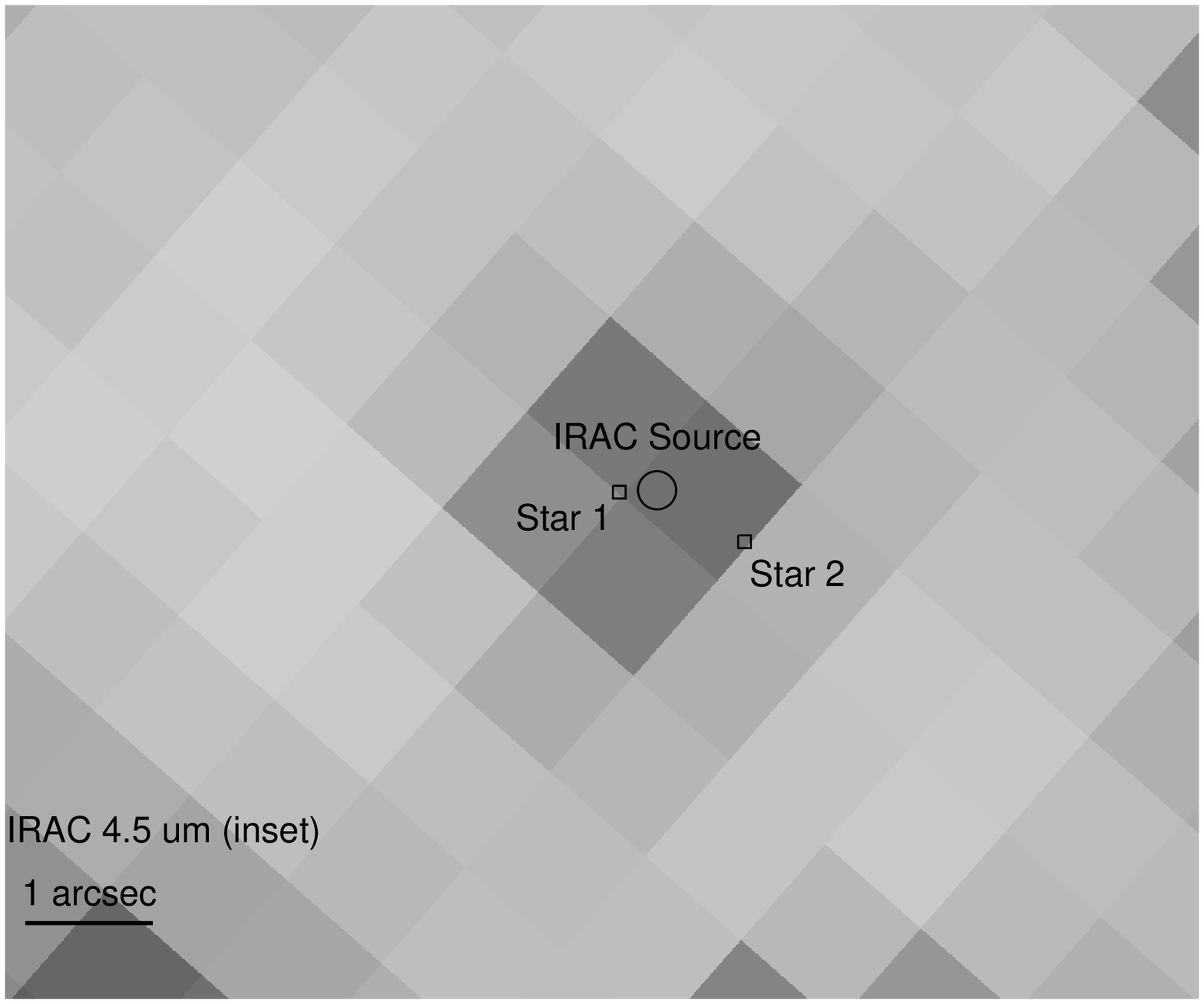}}
\caption{Field of \axp\ at $2.1\,\um$ (NIRI, {top left}),
  $4.5\,\um$ (IRAC, {bottom left}), and $8.0\,\um$ (IRAC,
  {top right}).  
On the IRAC images, we label several astrometric reference stars (A,
  D, and F) from \citet{htvk+01} and from this work (101 and 102),
  along with the IRAC source.  In the NIRI image, we label the same
  reference sources, plus we label stars 1 and 2 separately.  In the
  bottom right, we show a zoom around the position of \axp\ in the
  $4.5\,\um$ IRAC image (the region is indicated by the box in the
  larger image).  The position of the IRAC source is shown by the
  circle, with the circle's radius equal to the astrometric
  uncertainty ($0\farcs15$).  The positions of stars 1 and 2 from the
  NIRI image are also shown with appropriate error circles.  In all
  images, north is up, east is to the left, and scale bars are
  indicated in the lower left.}
\label{fig:images}
\end{figure*}

\section{Observations and Reduction}

\subsection{IRAC Data}
Our \dataset[ADS/Sa.Spitzer#23351296]{primary data} were observations
of \axp\ with the Infrared Array Camera (IRAC; \citealt{fha+04})
onboard the \textit{Spitzer Space Telescope} \citep{wrl+04} on
2007~August~8.  We observed \axp\ using two of the four possible IRAC
channels: $4.5\,\um$ and $8.0\,\um$ (IRAC channels 2 and 4).  The
observation consisted of 50 dithered exposures of $96.8\,$s for a
total integration of $80.7\,$minutes.

We started with the Basic Calibrated Data (BCD, from pipeline version
S16.1.0), discarding the first exposure at each wavelength as
recommended.  We then processed the BCD images though an artifact
mitigation pipeline\footnote{See
\url{http://spider.ipac.caltech.edu/staff/carey/irac\_artifacts/}.}
(2005~October~13 version).
With the \texttt{MOPEX} (MOsaicker and Point source EXtractor, ver.\
16.3.7) pipeline, we mosaiced the individual exposures together by interpolating the data onto a common grid and
rejecting radiation hits.
The final images are shown in Figure~\ref{fig:images}.

\subsection{Gemini Data}
Our analysis requires precise relative astrometry, and the angular
resolution of the IRAC images results in many blended objects.  To aid
in interpreting our mid-IR data and in particular to serve as an
improved astrometric reference, we analyzed an archival near-IR
(\ks-band, or $2.1\,\um$) observation taken with the Near-Infrared
Imager (NIRI; \citealt{hji+03}) on the 8\,m Gemini North telescope at
Mauna Kea, Hawaii.  The data consist of 50 exposures with the $f$/6
camera taken on 2003~May~27 (and previously published by
\citealt{tkvkd04}), each with $4\times15\,$s integrations, for a total
exposure of $50\,$minutes.  We used an \texttt{IRAF} package available
from the NIRI Web site\footnote{See
\url{http://www.us-gemini.noao.edu/sciops/instruments/niri/NIRIIndex.html}.}
to reduce the data.  We proceeded through the steps of this package,
flatfielding the data, subtracting the sky, shifting the images, and
adding them together.  We referenced the astrometry of the final image
to the Two Micron All Sky Survey (2MASS; \citealt{2mass}), finding 57
stars that were not saturated or badly blended, and getting rms
residuals of $0\farcs16$ in each coordinate.  This image is also shown
in Figure~\ref{fig:images}.

\section{Analysis}
We searched the IRAC images in Figure~\ref{fig:images} for a
counterpart to \axp\ by looking for a point source at the position
corresponding to star~1 (which is \axp; \citealt{htvk+01}) in the NIRI
image.  At this position, we see a faint point source in the IRAC
$4.5\,\um$ image (hereafter the ``IRAC source'').  However,
one must be careful, as the position of star 2 only differs from that
of the magnetar by $0\farcs97$; this is $\approx$1 IRAC pixel width,
and only half of the angular resolution of the image ($\approx 2\arcsec$
FWHM).

We performed photometry on the $4.5\,\um$ IRAC image to find the
positions and flux densities of all of the point sources.  We used the
\texttt{APEX} (Astronomical Point Source EXtraction, part of
\texttt{MOPEX}) software to identify and perform point-response
function (PRF) fitting for the photometry using the routine
\texttt{apex\_1frame.pl} with the \textit{Spitzer}-supplied PRF.  The
PRF is the traditional point-spread function (PSF) convolved with the
pixel-response function to determine how a source actually appears in
the data\footnote{The distinction between the PSF and PRF is important
for undersampled or critically sampled data like those here; see
\url{http://ssc.spitzer.caltech.edu/irac/psf.html}.}.
Part of this analysis identified blended objects using simultaneous
PRF fitting, but the IRAC counterpart was consistent with a single
point source (the $\chi^2$ per degree of freedom for the PRF fit was
$0.48$, consistent with other point sources in the field which have
reduced $\chi^2$ of 0.1--3) with flux
density $6.3\pm1.0\,\ujy$ at $4.5\,\um$.  The uncertainty here
includes the standard term from the PRF fitting but is dominated by
substantial contributions from difficulty in robustly measuring the
background and identifying which pixels make up the source.  At
$8.0\,\um$ there is no detection, and we estimate a limit of
$<20\,\ujy$ (95\% confidence, limited largely by the varying diffuse
background) based on the \texttt{APEX} detections in that region.

\begin{figure}
\plotone{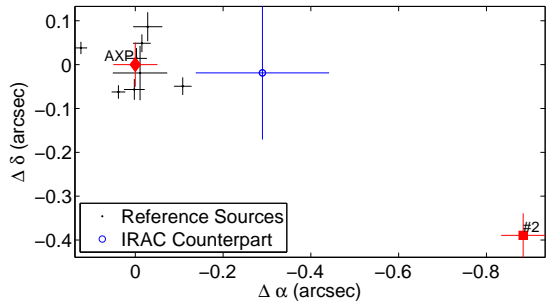}
\caption{Relative astrometry of the IRAC source.  We show the residual
position difference for eight reference sources (black points) that we
used to transform between the IRAC image and the much higher
resolution NIRI image after accounting for a net shift of $0\farcs1$.
We also show the measured position of the IRAC source (blue circle)
and the positions of star 1 (red diamond, labeled ``AXP'') and star 2
(red square).  The position of the IRAC source is much more consistent
with that of star 1 than star 2.  Also see the bottom right panel of Figure~\ref{fig:images}.}
\label{fig:astrom}
\end{figure}

In determining whether the IRAC source is indeed the counterpart to
star~1, we considered three issues.  First, we were concerned with
whether the position of the IRAC source is consistent with that of
star 1 or 2.  We took the astrometry (using both \texttt{APEX} and
standard centroiding; both gave consistent results) of the eight reference
sources (some of which are labeled in Figure~\ref{fig:images}), along
with the NIRI astrometry of those objects and stars 1 and 2.  We used
the reference stars to refine the astrometry of the IRAC image
relative to the NIRI image {(we did not use a proper reference
frame as an intermediary, as this would have introduced additional
uncertainties)}.  Fitting only for an offset (the position angles of
both observations were referenced independently to 2MASS), we found a
small shift ($0\farcs1$) with an rms of $0\farcs05$.  With this shift,
the position of the IRAC source differs from that of star 1 by
$0\farcs28$, and from that of star 2 by $0\farcs74$ (see
Figure~\ref{fig:astrom}).  There is an intrinsic centroiding
uncertainty for the IRAC source (the potential counterpart) of
$\approx 0\farcs15$ in each coordinate {(90\% confidence radius
of $0\farcs32$)}.  Given this, the IRAC source is largely consistent
with the position of star 1 (the probability of chance alignment is
78\%) and is inconsistent with the position of star 2 (the probability
of chance alignment is 99.998\%).  The situation is also illustrated
in the bottom right panel of Figure~\ref{fig:images}, which shows the
extent of the PRF of the IRAC counterpart, the measured centroid, as
well as the positions of stars 1 and 2 in the same reference frame.

\begin{figure}
\plotone{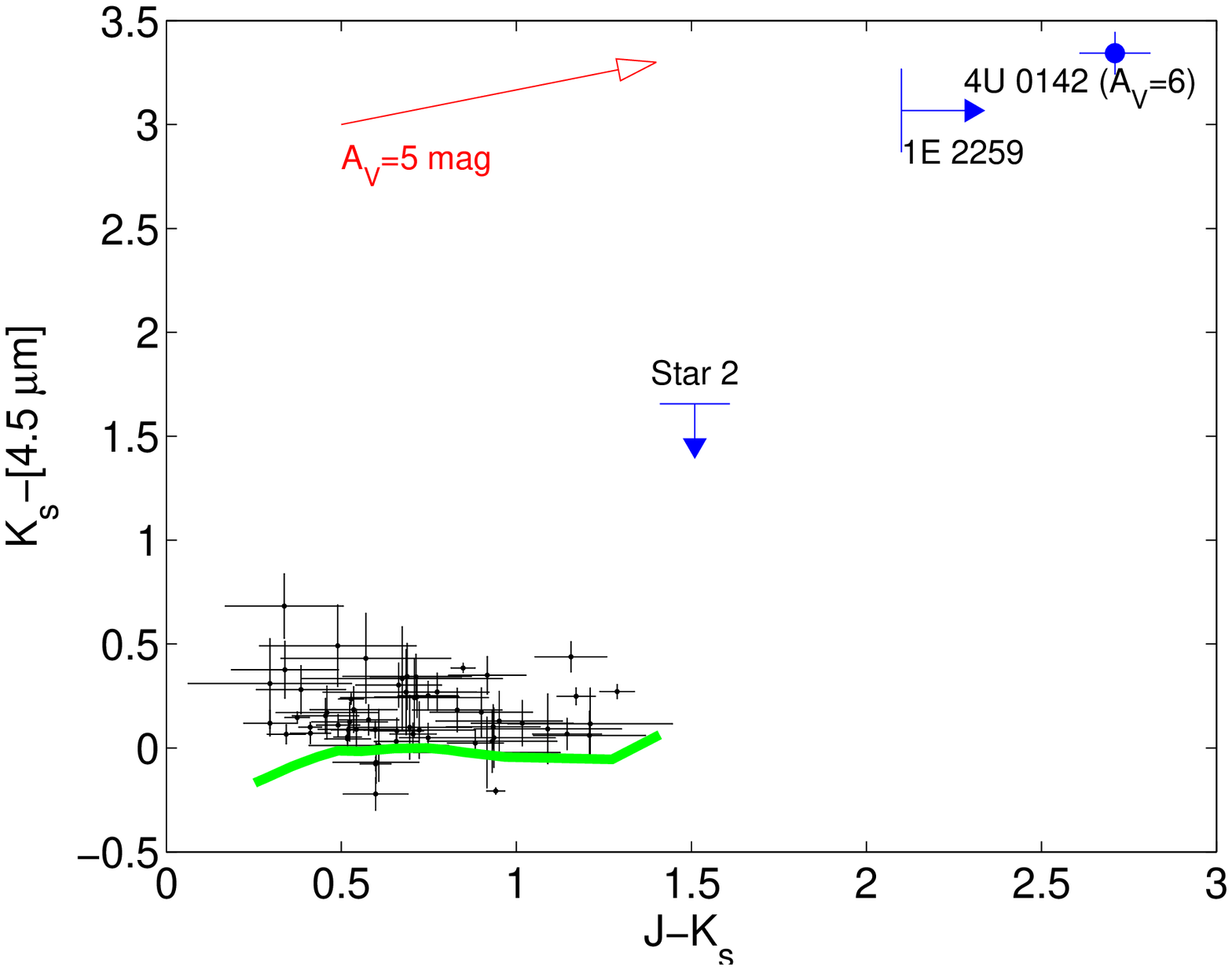}
\caption{$J-\ks$ color vs.\ $\ks-[4.5\,\um]$ color for the field of
  \axp.  The objects from the field with both 2MASS and IRAC
  detections are the black points.  \axp\ is the upper limit in the
  upper right, while star 2 is also an upper limit (this upper limit
  is only approximate, since star 2 is located within the PRF from
  \axp). We also plot the counterpart of 4U~0142+61 \citepalias{wck06}
  reddened from $A_V=3.5\,$mag \citep{dvk06b} to the same total
  extinction as \axp.    A reddening vector for $\av=5$ is shown, as is
  a main sequence with zero reddening (solid line, computed from
  \citealt{kurucz93} models; a giant branch will appear very similar).
  }
\label{fig:cc}
\end{figure}

Second, we compared the IR colors of the IRAC source to all of the
other objects in the field.  We show a color-color diagram in
Figure~\ref{fig:cc}.  For the IRAC source, we used the near-IR ($J$-
and \ks-band) photometry of star~1 given by \citet{htvk+01}; for star
2, we also used the photometry from \citet{htvk+01}; for the field
stars, we used 2MASS for the near-IR.  Assuming, as deduced from the
above astrometry, that the IRAC source is associated with star 1 and
not with star 2, we see that star 2 is consistent with the bulk of the
field population, i.e., reddened main-sequence stars.  On the other
hand, the IRAC source is significantly redder in both $J-\ks$ and
$\ks-[4.5\,\um]$ and clearly stands out from the population.  It is
quite close in colors to 4U~0142+61, for the same extinction.
Uncertainties in the visual extinctions of both \axp\ and 4U~0142+61
amount to $\pm2\,$mag \citep{dvk06b} and do not change the general
agreement because of the small extinctions at near- and mid-IR bands
($A_J\approx 0.3A_V$, $A_{K}\approx 0.1A_V$, $A_{4.5\,\um}\approx
0.05A_V$; also see Figure~\ref{fig:sed})).

Finally, we examined the spectral energy distributions (SEDs) of the
IRAC source and star 2 (Figure~\ref{fig:sed}).  First, we estimated the
spectral type of star 2.  With only two data points ($J=23.1\pm0.1$,
$\ks=21.5\pm0.2$) we could not discriminate between later stellar type
and higher reddening (the $R$- and $I$-band upper limits from
\citealt*{hvkk00} were not constraining), but we find reasonable solutions
for K/M dwarf stars with a few magnitudes of extinction and distances
of a few kpc (based on \citealt{allen}).  For stars of those spectral
types, we would expect the SED to peak in the near-IR region and
decline in the IRAC bands, with typical flux densities of $\approx
0.5\,\ujy$ expected.  Even brown dwarfs, which can be redder in
$\ks-[4.5\,\um]$, have $F_\nu(2.1\,\um)\gsim F_\nu(4.5\,\um)$
\citep{psb+06}. There are classes of objects such as young stellar
objects or massive stars with dusty winds that could produce such red
colors \citep[e.g.,][]{ukm+07,mll+07}, but they are relatively rare
{(although this is a complicated region with many objects along the
line of sight; \citealt{kuy02})}, often brighter than star 2 (spectral
types earlier than about F0V would have to be well outside the Galaxy
to have the observed $\ks$ magnitude), and would likely have been
identified at this position in other wavelengths (X-ray, near-IR,
molecular line, or continuum radio, etc.; see \citealt{kuy02,spg+04})
but were not.  In contrast, the $4.5\,\um$ flux density of the IRAC
source is a factor of 3 higher than the $\ks$-band flux densities of
star 1 or 2, and an order of magnitude higher than the expectation for
a normal star at $4.5\,\um$.

Taken together with the astrometry, this is strong support for not
associating the IRAC source with star 2, but instead with star 1.  It
is extremely unlikely that the IRAC source is associated with neither
object, as probability of chance coincidence with star~1 is only
0.3\%, and this would also imply even more extreme colors
(Figure~\ref{fig:cc}) that would be clearly non-stellar.  Even with
several tens of magnitudes of extinction it would be difficult to move
the IRAC source onto the stellar locus, and examining the NIRI image
we see no other sources with $\ks-[4.5\,\mu{\rm m}]>2$.  From
\citet{dvk06}, an extinction of only $\sim 10\,$mag is reached at
$\sim 8\,$kpc, and the molecular gas maps of Colden\footnote{See
\url{http://asc.harvard.edu/toolkit/colden.jsp}.} find a maximum
extinction of $\sim 5\,$mag, so having very high extinction outside of
very local extinctions seems unlikely.   We therefore conclude that the
IRAC source is indeed the mid-IR counterpart to star~1/\axp.

\section{Discussion}
\begin{figure}
\plotone{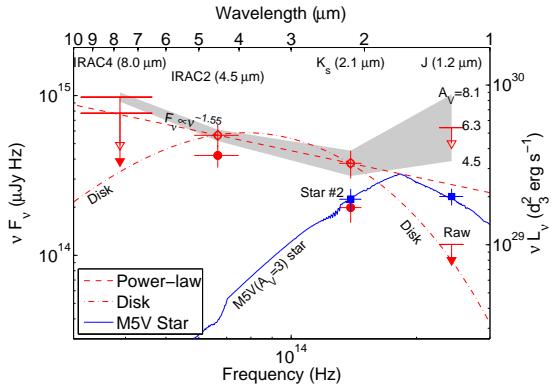}
\caption{SED of \axp\ and star 2.  We plot
  $\nu F_\nu$ (left axis) vs.\ frequency for \axp\ (solid red
  points/upper limits) and star 2 (blue squares), and also give the
  luminosity $\nu L_\nu$ (right axis) for a distance of $3d_3\,$kpc.
  For star 2 we also show one of the possible stellar fits (an M5V
  star at $\av=3$).  The open red points/upper limits are the data on
  \axp\ dereddened with $\av=6.3$ using the reddening law of
  \citet{imb+05}, and the shaded band shows the $\pm 1\,\sigma$ range
  of $A_V$ \citep{dvk06b}.  We plot a power law, $F_{\nu}\propto
  \nu^{-1.55}$, which fits both detections for \axp\ (dashed line)
  and an irradiated disk model (based on \citetalias{wck06};
  dot-dashed line).  The various bands are labeled.}
\label{fig:sed}
\end{figure}

We have found a $4.5\,\um$ counterpart to the magnetar \axp, with flux
density $6.3\pm1.0\,\ujy$ and a limit of $<20\,\ujy$ at $8.0\,\um$.
The absorption-corrected mid-IR/X-ray\footnote{The unabsorbed X-ray
flux of \axp\ is $\approx3\times10^{-11}\,{\rm erg}\,{\rm
s}^{-1}\,{\rm cm}^{-2}$ \citep{pkw+01}, although this is only over the
2--10$\,$keV range and does not include softer emission which is
uncertain due to the moderate absorption.} flux ratio for this source
is $1.9\times 10^{-4}$, very similar to the ratio for 4U~0142+61, so
the common X-ray to $2.1\,\um$ flux ratios \citep{dvk05} seem to
extend further into the infrared.  We note, though, that one must be
careful with {\em any} model or interpretation for the IR emission
since the $2.1\,\um$ flux density of \axp\ (and of 4U~0142+61 for that
matter; \citealt*{hvkk04}; \citealt{dvk06c}) is known to vary.  The
$2.1\,\um$ flux density we plot in Figure~\ref{fig:sed} is the
faintest measured and is assumed to be close to the baseline level
\citep{tkvkd04} and should not be affected by flares, but it was
measured two years before the IRAC observations.  {For \axp, the IR
variability had been seen along with X-ray flaring \citep{kg02,kgw+03}
in the past. Since then, though,} no transient X-ray behavior for
\axp\ was reported and the X-ray flux is now near the quiescent level
(although with a slightly different spectrum; \citealt{zkd+08}), so
{our assumption of a baseline flux seems safe}, but X-ray monitoring
is not very regular{ and there could potentially be IR variability
independent of X-ray activity.}

{
With the limited data we have, a wide range of model fits are
possible.  Here we discuss two general categories: a power law and an
irradiated disk model.}  
A simple power law ($F_\nu\propto\nu^{-1.55}$) can fit the two
detected IR flux points and is also below the upper limits at
$1.2\,\um$ and $8\,\um$ (Figure~\ref{fig:sed}).  Such a power law
spectrum could arise from the pulsar magnetosphere \citep[as for Crab
pulsar; see][]{tgw+06}, which could also possibly produce correlated
X-ray and infrared flux changes, although this has not been studied in
detail.  However, unlike the Crab pulsar, this power-law rises further
into the mid-IR band, and it would be interesting to obtain
longer-wavelength data on \axp\ to see if the SED keeps rising past
$4.5\,\um$, but source confusion makes this very difficult. The rising
spectrum is problematic for some detailed magnetospheric models (like
those of \citealt{ec04} and \citealt{hh05}), but a generic
magnetospheric origin is certainly possible.

However, as with 4U~0142+61 \citepalias{wck06}, we can also fit the
1E~2259+586 data with an irradiated passive disk model (also see
\citealt{vrg+90,phn00}; other spectral shapes are also possible, but
the fits are too unconstrained and without specific motivation we will
not address them).  
{Without more data, we cannot make conclusions about the
presence or absence of a disk, and indeed other observations suggest
that a disk interpretation is problematic \citep{wbk+08}.  However, it
is still possible, and we find it worthwhile to expand on the
discussion of \citetalias{wck06} and further explore the implications
such a disk would have, noting in particular where \axp\ differs from
4U~0142+61.} 

This model fit is qualitatively similar to what we
obtained in \citetalias{wck06} for 4U~0142+61, with a small inner
radius and an outer radius a factor of a few to 10 larger.  The
details of both fits depend on assumptions about the distances,
extinctions, and inclinations, but the rough shapes of the SEDs going
from the near- to mid-IR are similar, with a factor of 3 in flux
density increase from $2.1$ to $4.5\,\um$, after correcting for
extinction (unlike what \citealt{wbk+08} found for 1E~1048.1$-$5937).
 The inner radius is constrained by the need to have the
$\ks$-band measurement lie below the $4.5\,\um$ measurement, while the
outer radius is less constrained by the $8.0\,\um$ upper limit.  Note
that the nominal inner radius of $R_{\rm in}=0.25R_{\odot}$ (for a
distance of 3\,kpc and an inclination of $60\degr$) is smaller than
the light cylinder radius ($0.4R_{\odot}$) which would imply
interaction between the disk and the magnetar's spin
\citep[e.g.,][]{chn00}---possibly contradicted by its observed steady
spin-down \citep{gk02}.  This could either point to an inconsistency
in our model, or simply that we must choose the other model parameters
(distance, inclination, etc.)  such that the inner radius is $\gsim
0.5R_{\odot}$; as an example, a larger distance such as that
determined by \citet{dvk06} has an inner radius\footnote{As mentioned
in \citetalias{wck06}, the inner radius discussed here is just that to
which dust can penetrate.  Gas could extend further inward, so that
even with the inner dust radius larger than the light cylinder radius
interactions could still occur.}  more than twice the light cylinder
radius.

The binding energy of the putative disk, following \citetalias{wck06}
to estimate the disk mass, is $\sim10^{45}\,{\rm erg}$ for the upper
end of the mass estimate.  This is large enough that we do not have to
worry that such a disk would have been unbound by the large X-ray
bursts (a total fluence of $\expnt{3}{-8}\,{\rm erg}\,{\rm s}^{-1}$, or
total isotropic energy of $\expnt{3}{37}d_3^2{\rm\, erg}$) found by
\citet{kgw+03}, and the physical state and structure of the disk may
not be altered dramatically, as this energy is equivalent to only
$\sim10^3\,$s of normal X-ray activity.  Even a disk several order of
magnitude lower in mass would remain bound.  Only a truly giant flare,
such as those seen from several soft $\gamma$-ray repeaters
\citep{wt06}, would be energetic enough to disrupt a disk like this,
although the details of that process are difficult (also see
\citealt{wrrd+08}), but we stress that our mass estimate is an upper
limit, and at the low end of the allowed mass range the disk would not
be bound.  {Again, this could indicate either a problem with the disk 
model, or that objects with disks cannot have had large flares.}

Among magnetars, the correlated near-IR and X-ray variability found by
\citet{tkvkd04} for \axp\ is unique.  Such a correlation is a natural
(but not necessarily unique) consequence of the disk model: a change
in the X-ray flux would produce an accompanying change in the infrared
flux from the disk, and while the total reprocessed flux is a constant
fraction of the X-ray flux, the IR flux at a given wavelength can
change faster or slower as annuli of a given temperature move and
change area
\citep[cf.][]{vpm94}.  We find that the $2.1\,\um$ flux increases
seen by \citet{tkvkd04}, which were of comparable amplitude to the
X-ray flux increases, are roughly consistent with a disk like that
considered here, although this assumes that 100\% of the $2.1\,\um$
flux comes from the disk and ignores the possibility of physical
changes to the disk such as movement of the inner radius following the
flux changes.  However, correlated changes were not seen in 4U~0142+61
\citep{dvk06c}, who instead saw relatively rapid, significant changes
at $2.1\,\um$ without any changes in X-rays.  Any model must be able
to accommodate this wide range in behavior (see \citealt{wk08}).
Additional observations at longer wavelengths where the decomposition
is less ambiguous could again provide important constraints.

\acknowledgements We thank an anonymous referee for helpful comments.
This work is based on observations made with the \textit{Spitzer Space
Telescope}, which is operated by the Jet Propulsion Laboratory,
California Institute of Technology under a contract with NASA. Support
for this work was provided by NASA through an award issued by
JPL/Caltech.  Support for D.L.K.\ was provided by NASA through Hubble
Fellowship grant 01207.01-A awarded by the Space Telescope Science
Institute, which is operated by the Association of Universities for
Research in Astronomy, Inc., for NASA, under contract NAS 5-26555 and
by the National Science Foundation under Grant PHY05-51164.  Guest
User, Canadian Astronomy Data Centre, which is operated by the
Herzberg Institute of Astrophysics, National Research Council of
Canada.

{\it Facilities:} \facility{Gemini:North (NIRI)}, \facility{\textit{Spitzer} (IRAC)}

\end{document}